\documentclass[aps,prl,twocolumn,showpacs,groupedaddress,preprintnumbers]{revtex4-1}  

\usepackage{graphicx}  
\usepackage[caption=false]{subfig}
\usepackage{dcolumn}   
\usepackage{bm}        
\usepackage{amssymb}   
\usepackage{braket}
\usepackage{amsmath}
\usepackage{color}  	

\newcommand{\bT}{\mathbf{T}}
\newcommand{\beq}{\begin{equation}}
\newcommand{\eeq}{\end{equation}}
\newcommand{\bea}{\begin{eqnarray}}
\newcommand{\eea}{\end{eqnarray}}
\newcommand{\dd}{\text{d}}

\newcommand{\bd}[1]{\mathbf{#1}}

\newcommand{\rbar}{\bar{\rho}}
\newcommand{\sbar}{\bar{\sigma}}
\newcommand{\tbar}{\bar{\tau}}
\newcommand{\dbra}[1]{[ #1|}
\newcommand{\dket}[1]{|#1]}

\begin{document}

\preprint{MAN/HEP/2020/008}
\preprint{UWTHPH-2019-10}
\preprint{MCnet-20-11}
\title{Resummation and simulation of soft gluon effects beyond leading colour}
\author{Matthew De Angelis}
\affiliation{Consortium for Fundamental Physics, School of Physics \&
	Astronomy, University of Manchester, Manchester M13 9PL, United
	Kingdom}
\author{Jeffrey R. Forshaw}
\affiliation{Consortium for Fundamental Physics, School of Physics \&
	Astronomy, University of Manchester, Manchester M13 9PL, United
	Kingdom}
	\affiliation{Erwin Schr\"odinger Institute for Mathematics and Physics, University of Vienna, 1090 Wien, Austria}
\author{Simon Pl\"atzer}
	\affiliation{Particle Physics, Faculty of Physics,
University of Vienna, 1090 Wien, Austria}
\affiliation{Erwin Schr\"odinger Institute for Mathematics and Physics, University of Vienna, 1090 Wien, Austria}
\date{\today}

\begin{abstract}
We present first results of resumming soft gluon effects in a
simulation of high energy collisions beyond the leading-colour
approximation. We work to all orders in QCD perturbation theory using
a new parton branching algorithm. This amplitude evolution algorithm
resembles a parton shower that is able to systematically include
colour-suppressed terms. We find that colour suppressed terms can
significantly contribute to jet veto cross sections.
\end{abstract}

\maketitle

{\it Introduction} -- We present first results from a new Monte Carlo
code, CVolver, that simulates high-energy particle
collisions \footnote{CVolver stands for Colour Virtual Evolver, and
  originates from the studies first presented in
  \cite{Platzer:2013fha}}. Our analysis here is based upon simulated
two-jet events and our goal is to improve on the accuracy of existing
simulations
\cite{Bahr:2008pv,Bellm:2015jjp,Sjostrand:2014zea,Bothmann:2019yzt} by
including colour correlations beyond the leading-colour
approximation. To do so we pursue a new paradigm of evolution at the
amplitude level in place of the traditional probabilistic algorithms
(see also
\cite{Nagy:2012bt,Nagy:2015hwa,Martinez:2018ffw,Forshaw:2019ver,Nagy:2019bsj,Nagy:2019pjp}). In
the next section we introduce the theoretical framework, with
particular emphasis on our treatment of colour. In the following
section, we present results for the cross section for the production
of a two-jet system with a restriction on the amount of radiation
lying in some angular region outside of the jets. This process is
sensitive to wide-angle, soft-gluon emission and thus provides a good
test of the framework. We observe significant deviations from the
leading-colour approximation.
\vspace*{1ex}

{\it Summing soft-gluon effects} -- Particle collisions involving
coloured particles and a large transfer of momentum, $Q$, are
computable using perturbative QCD. However, fixed-order perturbation
theory is often insufficient due to the presence of large logarithms
that compensate the smallness of the perturbative coupling,
$\alpha_s$. In other words, there exist terms of order $\alpha_s^n
L^m$ where $L$ is some large logarithm and $m \le 2n$. Large
logarithms can arise if gluons are emitted with a low energy compared
to the large momentum transfer, and if the observable is sensitive to
those emissions. We refer to these as soft-gluon logarithms and, in
\cite{Martinez:2018ffw}, we presented an iterative algorithm for
summing them to all orders in perturbation theory for general
short-distance scattering processes. In this paper, we will sum the
most important of these logarithms, though the framework is general
enough to go beyond this `leading logarithmic approximation'. The
formalism can be extended to also include logarithms of collinear
origin \cite{Forshaw:2019ver,Forshaw:2020wrq}.  The differential cross
section for $n$ soft-gluon emissions can be written
\begin{align}
\dd \sigma_n & = \text{Tr} \, \mathbf{A}_n \, \dd \Pi_n~, \label{eq:sign}
\end{align}
where the operators $\mathbf{A}_n$ satisfy the recurrence relation
\begin{eqnarray}
\mathbf{A}_n(E) = \mathbf{V}_{E,E_n}\mathbf{D}_n^\mu \; \mathbf{A}_{n-1}(E_n)
\; \mathbf{D}_{n\mu}^\dag \mathbf{V}_{E,E_n}^\dag \, \Theta(E \le E_n), \nonumber \\ \label{eq:recur}
\end{eqnarray}
where $\Theta(E \le E_n)$ is the Heaviside function. The virtual-gluon
(Sudakov) operator sums over all single-gluon exchanges between pairs
of partons. For two-jet production off a colour singlet (e.g. $V \to q
\bar{q}$ or $H \to gg$) it is given by \footnote{For processes with
  two or more coloured particles in both the initial and final state
  of the hard process, Coulomb/Glauber exchanges are also needed.}
\bea {\mathbf{V}}_{a,b} &=& \exp \Bigg[ -\frac{\alpha_s}{\pi} \ln
  \left(\frac{b}{a}\right) \; \sum_{i<j} (-\mathbf{T}_i \cdot
  \mathbf{T}_j) \, \nonumber \\ & & \times \int \frac{\dd \Omega}{4\pi
  } \frac{n_i \cdot n_j}{(n_i \cdot n) \; (n_j \cdot n)} \Bigg] ~,
\nonumber \\ \label{eq:simple} \eea where $n_i$ is a light-like vector
whose spatial part is the unit vector in the direction that particle
$i$ travels.  The integral is over the direction of the light-like
vector $n$ and, for now, we take $\alpha_s$ to be fixed. The real
emission operator and phase-space factor are
\begin{align} \mathbf{D}_i^\mu &= \sum_j \mathbf{T}_j \; E_i
\frac{p_j^\mu}{p_j \cdot q_i} ~~~\text{and} \nonumber \\ \dd \Pi_n &=
\prod_{i=1}^n \left(-\frac{\alpha_s}{\pi} \frac{\dd E_{i}}{E_{i}}
\frac{\dd \Omega_{i}}{4\pi } \right) ~. \label{eq:simple1}
\end{align}
Note that the sum over partons in the definitions of
$\mathbf{V}_{a,b}$ and $\mathbf{D}_i^\mu$ is context-specific, i.e. it
runs over all prior soft-gluon emissions in addition to the partons in
the original hard scattering. The colour charge operator,
$\mathbf{T}_j$, is also in a context-specific representation of
SU(3)$_c$. Soft-gluon evolution proceeds iteratively starting from a
hard-scattering operator, $\mathbf{H} = |\mathcal{M}\rangle \langle
\mathcal{M}|$ with $\mathbf{A}_0(E) = \mathbf{V}_{E,Q} \, \mathbf{H}
\, \mathbf{V}_{E,Q}^\dag$.  A general observable, $\Sigma$, can be
computed using
\begin{eqnarray}
\Sigma(\mu) &=& \int \sum_n ~ \dd \sigma_n \,
u_n(k_1,k_2,\cdots,k_n)~, \label{eq:eord}
\end{eqnarray}
where the $u_n$ are the observable dependent measurement functions and
the $k_i$ are soft-gluon momenta. We suppress the dependence on the
hard partons and integration over their phase space. Although we
assume energy ordering, this is not essential and the algorithm can
readily be adapted to account for a different ordering variable.  We
should take the limit $\mu \to 0$ in Eq.~(\ref{eq:eord}), though it is
also correct to put $\mu = Q_0$ if the observable is fully inclusive
over gluon emissions with $E < Q_0$.

\begin{figure}[t]
	\includegraphics[width=0.49\textwidth]{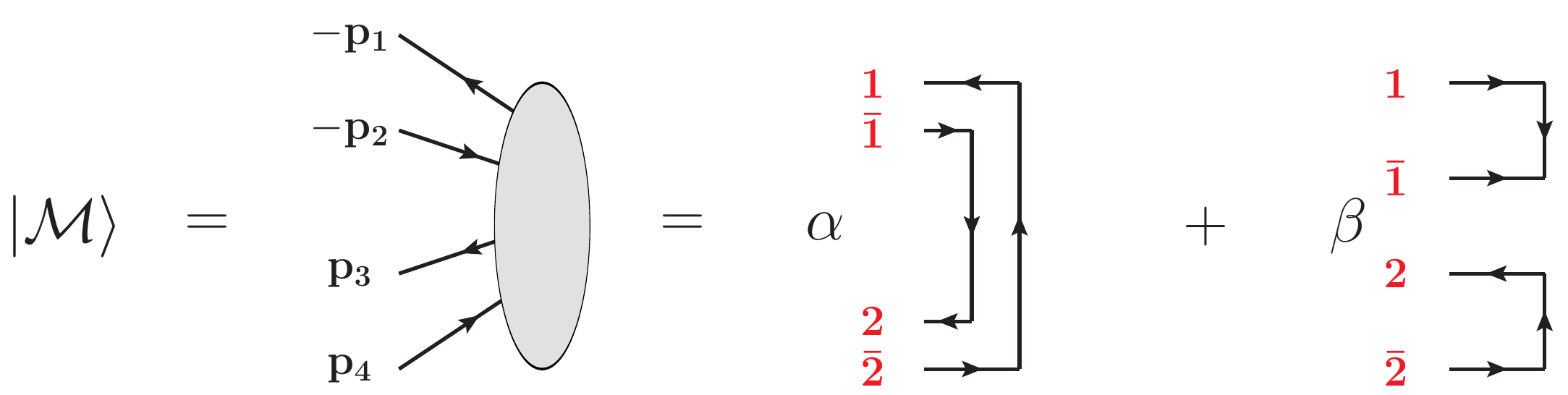}
	\caption{\label{fig:flow} An example of an amplitude
          decomposed in the colour flow basis.}
\end{figure}

\begin{figure}[t]
	\centering
	\includegraphics[width=0.49\textwidth]{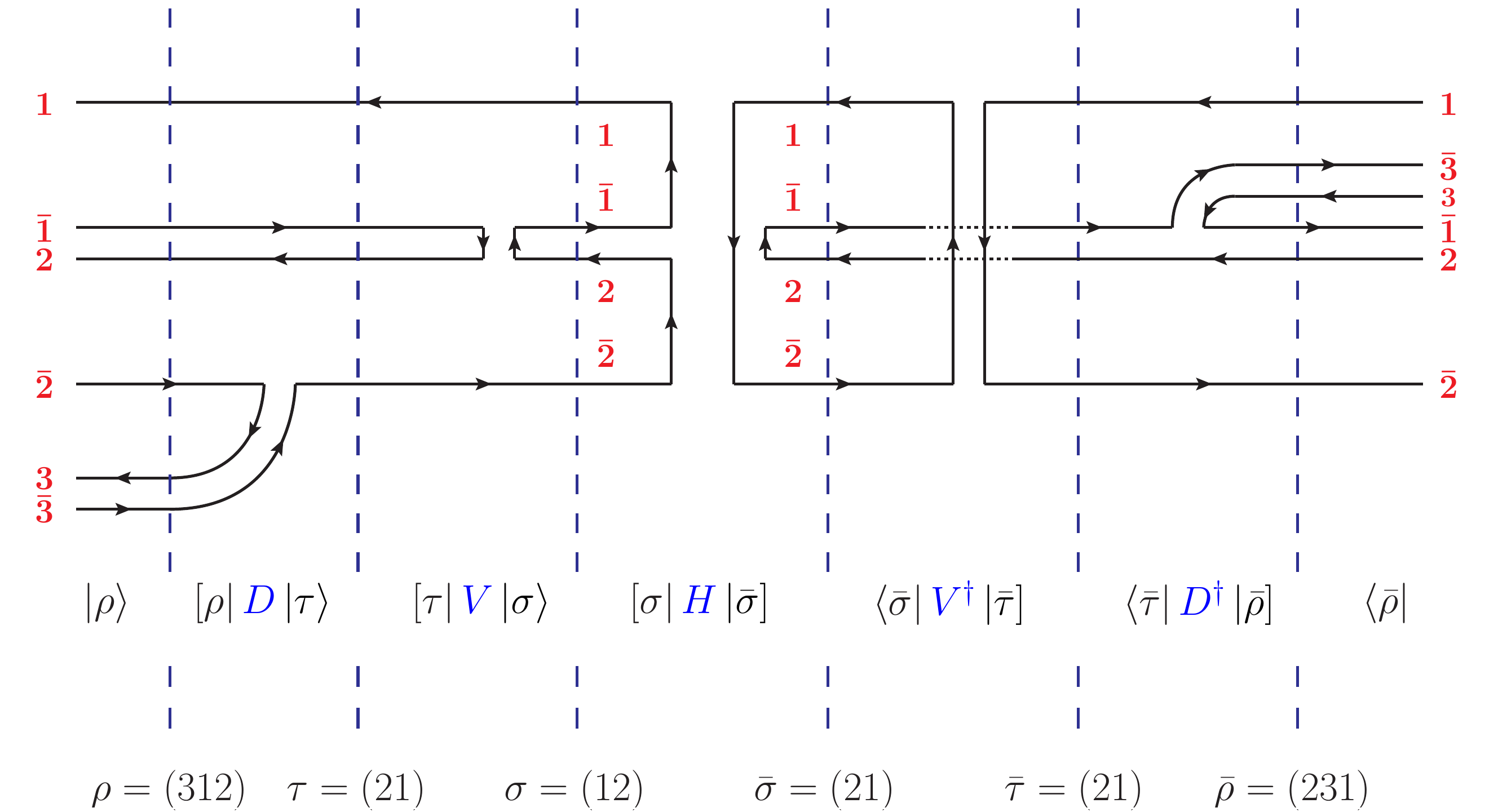} \\
	\caption{One contribution to the $\mathbf{A}_1$ operator,
          starting from the hard scattering illustrated in
          Fig.~\ref{fig:flow}. It corresponds to single gluon emission
          with two virtual gluon exchanges. The vertical dotted lines
          are to help identify the intermediate colour states. The
          algorithm works iteratively outwards, starting from the hard
          process in the middle and multiplying matrix elements as it
          goes.}
	\label{fig:A}
\end{figure}

This iterative form of the algorithm is well suited to a Monte Carlo
implementation. The kinematic part of the evolution is diagonal and
does not pose any new problems. The main challenge is to account for
the independent colour evolution in the amplitude and the conjugate
amplitude. To do this we use the colour-flow basis, which is
illustrated in the case where the hard scattering process is $q
\bar{q} \to q \bar{q}$ in Fig.~\ref{fig:flow}. We choose to draw all
particles in the amplitude as heading off to the left and negative
momenta indicate incoming particles. The particle with incoming
momentum $p_1$ is an anti-quark, that with incoming momentum $p_2$ is
a quark whilst $p_3$ and $p_4$ label an outgoing quark and an outgoing
anti-quark respectively. In the colour-flow basis, quarks and
anti-quarks are represented by colour and anti-colour lines (an
incoming quark is represented by an anti-colour line), whilst gluons
are represented by a pair of colour and anti-colour lines. A basis
vector in the colour space is then represented by stating how the
colour and anti-colour lines are connected. We see this on the
right-hand side of Fig.~\ref{fig:flow}. In the first term, the
incoming anti-quark is colour connected to the outgoing anti-quark and
the incoming quark is colour connected to the outgoing quark. The
second term is the only other possibility for two colour lines and
corresponds to the case where the incoming quark is colour connected
to the incoming anti-quark. Our convention is to write amplitudes such
that Fig.~\ref{fig:flow} corresponds to \beq \ket{\mathcal{M}} =
\alpha \ket{2 1} + \beta \ket{1 2}~.  \eeq A general state consisting
of $n$ colour lines has a basis of dimension $n!$ corresponding to all
possible permutations of the numbers $(1,2,\cdots,n)$. We normalize
the basis vectors so that \beq \braket{\alpha|\beta} = N_c^{n -
  \#(\alpha,\beta)}, \eeq where $\#(\alpha,\beta)$ is the minimum
number of pairwise swaps by which the permutations $\alpha$ and
$\beta$ differ, e.g. $\braket{12|12} = N_c^2$ and $\braket{12|21} =
N_c$. This basis is over-complete and not orthogonal but is very
simple to implement and provides excellent opportunities for
importance sampling. We introduce a dual basis, $|\alpha]$, such that
  $\bra{\alpha} \beta ] = [ \alpha \ket{\beta} =
      \delta_{\alpha\beta}$, where $\delta_{\alpha\beta}$ is unity if
      the two permutations are equal and zero otherwise. Also,
      $\sum_\alpha \ket{\alpha} [\alpha| = \mathbf{1}$. The trace in
        Eq.~(\ref{eq:sign}) is then computed using \beq \text{Tr} \,
        \mathbf{A}_n = \sum_{\sigma,\tau} [ \tau | \mathbf{A}_n |
          \sigma] \braket{\sigma|\tau} ~.  \eeq Fig.~\ref{fig:A}
        illustrates how we Monte Carlo over intermediate colour states
        by inserting the unit operator between successive real
        emission and virtual correction operators.
We select initial colour flows, $\sigma$ and $\bar{\sigma}$, and
compute the corresponding hard scattering matrix $[\sigma | \mathbf{H}
  | \bar{\sigma}]$. Then we choose the momentum of the first real
emission and, in anticipation of inserting Sudakov operators that act
on the amplitude and the conjugate (they evolve from the hard scale
$Q$ down to the energy of the emitted gluon, $E_1$), we choose two new
colour flows, $\tau$ and $\bar{\tau}$, from the set of all possible
colour flows that can be accessed after the action of a Sudakov
operator (see below). We then multiply the original hard scattering
matrix element by the corresponding Sudakov matrix elements. The
colour states $\rho$ and $\bar{\rho}$ are chosen next, from the set of
all possible permutations accessible after the emission of a gluon and
these are used to deduce the corresponding real emission matrix
elements. This whole process repeats until the evolution terminates
and the final product of matrix elements must be further multiplied by
the scalar product matrix, $\braket{\sigma_m | \bar{\sigma}_m}$, where
$m$ labels the final colour flows. This induces a $1/N_c$ suppression
factor for every swap by which the final permutations $\sigma_m$ and
$\bar{\sigma}_m$ differ.

Eq.~\eqref{eq:recur} can be re-written explicitly in terms of matrix
elements such that one step in the evolution is determined by
\begin{align} \label{eq:full}
M_{\rho \rbar}(E) = -\frac{\alpha_s}{\pi} & \frac{\dd E}{E} \frac{\dd
  \Omega}{4 \pi} \, \sum_{\substack{\tau, \sigma \\ \bar{\tau},
    \bar{\sigma}}} \dbra{\rho}\bd{D}_E \ket{\tau} \dbra{\tau}
\bd{V}_{E,E'} \ket{\sigma} \nonumber \\ & M_{\sigma\sbar}(E')
\bra{\sbar}\bd{V}^\dag_{E,E'} \dket{\bar{\tau}} \bra{\tbar}
\bd{D}_E^\dag \dket{\rbar} ,
\end{align}
where $E$ is the energy of the latest emission and $E'$ is the
previous energy. This expression is the core of our implementation and
it provides a map from a pair of colour flows $(\sigma, \sbar)$ to the
pair $(\rho,\rbar)$.

Calculating the effect of real emission matrix elements and virtual
gluon corrections involves computing matrix elements $[\tau |
  \mathbf{T}_i \ket{\sigma}$ and $[\tau | \mathbf{T}_i \cdot
    \mathbf{T}_j \ket{\sigma}$. Explicit expressions for these are
    presented in \cite{Martinez:2018ffw}. For the real emissions, the
    calculations are simplified since the real emission operator can
    either: (a) add a new colour line without changing any of the
    existing colour connections or (b) add a new colour line and then
    make a single swap. In the case that the gluon is emitted off a
    colour line this swap connects the colour of the emitted gluon to
    the anti-colour partner of the emitter (and likewise if the gluon
    is emitted off an anti-colour line). Taken together, a real
    emission in the amplitude and the conjugate amplitude only changes
    the number of swaps by which the two colour flows (in the
    amplitude and conjugate) differ by at most two. Specifically, $ 0
    \le \#(\sigma_{n+1},\tau_{n+1}) - \#(\sigma_{n},\tau_{n}) \le
    2$. This means that real emissions never bring the two colour
    flows `closer together'. `Singlet' gluons, as emitted in case (a),
    are subleading in colour and inert from the point of view of the
    subsequent evolution.  The simplicity of the real emissions means
    we do not need to make any approximation when computing them.

The main challenge is to compute the Sudakov matrix elements, $[\tau |
  \mathbf{V} \ket{\sigma}$, which involves the exponentiation of a
  possibly large colour matrix. The evaluation can be considerably
  simplified if we are prepared to sum terms accurate only to order
  $1/N_c^d$, where $d$ is a positive integer, while keeping the
  leading diagonal terms proportional to $(\alpha_s N_c)^n$ to all
  orders $n$. Choosing larger values of $d$ will lead to more accurate
  results that take a longer time to compute. To do this we use the
  result presented in \cite{Platzer:2013fha}: \bea [\tau | \mathbf{V}
    \ket{\sigma} & \approx & \delta_{\tau \sigma} R(\{ \sigma \}) +
    \sum_{l=1}^d \left( -\frac{1}{N_c} \right)^l \sum_{\sigma_0,
      \sigma_1,\cdots \sigma_{l}} \delta_{\tau \sigma_0}
    \delta_{\sigma_{l} \sigma} \nonumber \\ & & \times \prod_{\alpha
      =0}^{l-1} \Sigma_{\sigma_\alpha \sigma_{\alpha+1}} \, R(\{
    \sigma_0, \sigma_1, \cdots, \sigma_{l}\}) ~.  \label{eq:simon}
    \eea The anomalous dimension matrix elements are \beq [\tau | \ln
      \mathbf{V} \ket{\sigma} = \left( - N_c \, \Gamma_\sigma +
      \frac{\rho}{N_c} \right) \delta_{\tau \sigma} +
      \Sigma_{\tau\sigma}~. \label{eq:lnV} \eeq Explicit expressions
      for $\Gamma_\sigma$, $\rho$ and $\Sigma_{\sigma \tau}$ can be
      computed using the results presented in
      \cite{Platzer:2013fha,Martinez:2018ffw}. Note that the
      off-diagonal contribution, $\Sigma_{\tau\sigma}$, is only
      non-zero if \mbox{$\#(\sigma,\tau) = 1$}. In other words, each
      virtual gluon exchange can either leave a colour flow unchanged
      or induce a single swap. Exponentiating this to produce the
      Sudakov operator generates the possibility of many swaps but
      this can be managed since each swap comes at the price of a
      factor $1/N_c$. This is what allows us to truncate the sum in
      Eq.~(\ref{eq:simon}) at a small value of $d$. We refer to $d=0$
      as our leading colour virtuals (LC$_V$) approximation and $d=1$
      as next-to-leading (NLC${}_V$) etc. The kinematic functions $R$
      appearing in Eq.~(\ref{eq:simon}) are also listed in
      \cite{Platzer:2013fha}. In the case $d=0$, we only need
      \beq \label{eq:R1} R(\{\sigma\}) = e^{ - N_c \Gamma_\sigma'}~,
      \eeq where $\Gamma'_\sigma = \Gamma_\sigma -
      \rho/N_c^2$. Keeping only this term generates the leading colour
      approximation, improved by summing the colour-diagonal
      subleading parts \footnote{The improvement is only relevant if
        there are quarks in the hard process since $\rho$ always
        vanishes for purely gluonic states}. For $d=1$ we also need
\begin{align} \label{eq:R2}
R(\{\tau,\sigma\}) & = \frac{e^{-N_c \Gamma'_\tau} - e^{-N_c
    \Gamma'_\sigma}}{\Gamma'_\tau - \Gamma'_\sigma}
\end{align}
so that the NLC${}_V$ Sudakov matrix elements are \beq [\tau |
  \mathbf{V} \ket{\sigma} = \delta_{\tau \sigma} e^{-N_c
    \Gamma'_\sigma} - \frac{1}{N_c} \Sigma_{\tau \sigma} R(\{\tau,
  \sigma\}).  \eeq Notice that our N$^d$LC${}_V$ approximation
  involves at most $d$ swaps for each application of the Sudakov
  operator, which makes the task of Monte Carloing over accessible
  colour states tractable. It should be emphasised that since we treat
  the real emissions, the scalar product matrix and the diagonal part
  of the anomalous dimension matrix without any approximation, our
  leading-colour approximation goes well beyond summing only the
  strictly leading ($m=0$) terms in an expansion in $(N_c \alpha_s)^n
  / N_c^m$, see also \cite{Holguin:2020oui} for a recent discussion.

One stringent test of the algorithm is that it should correctly handle
the cancellation of collinear singular terms. We deal with the
collinear region by cutting out small cones around every real
emission. Specifically, we impose that $n \cdot n_{i,j} > \lambda$ for
emission off the $ij$ pair.  Correspondingly, for the loop integrals
we regulate using the replacement
\begin{align}
\frac{n_i \cdot n_j}{n_i \cdot n \, n_j \cdot n} & \to \\\nonumber
& \hspace{-1cm} \frac{n_i \cdot n_j}{n\cdot n_i + n \cdot n_j} \left(
\frac{\Theta(n \cdot n_i - \lambda)}{n \cdot n_i} + \frac{\Theta(n
  \cdot n_j - \lambda)}{n \cdot n_j} \right)~.
\end{align}
After integrating over solid angle, 
\begin{align}\label{eq:cool}
\int \frac{\dd \Omega}{4\pi} \frac{n_i \cdot n_j}{n\cdot n_i + n \cdot
  n_j} & \left( \frac{\Theta(n \cdot n_i - \lambda)}{n \cdot n_i} +
\frac{\Theta(n \cdot n_j - \lambda)}{n \cdot n_j} \right) \\\nonumber
& \approx \ln \frac{n_i \cdot n_j}{\lambda} ~,
\end{align}
and the only approximation is to disregard terms suppressed by powers
of the cutoff $\lambda$.  Since the $\ln \lambda$ term in
\eqref{eq:cool} is independent of $i$ and $j$ we could exploit colour
conservation to write
\begin{equation}
\sum_{i<j}
(-\mathbf{T}_i \cdot \mathbf{T}_j) = \frac{1}{2}\sum_{i} \mathbf{T}_i^2~, 
\end{equation}
which is colour diagonal. This leads to the well-known result that the
collinear region has trivial colour and it means that all of the
collinear cutoff dependence is in the abelian sector
\cite{Forshaw:2019ver}.

To perform the evolution efficiently, we use the Sudakov veto
algorithm with competition \cite{Platzer:2011dq,Kleiss:2016esx} in
order to select the energy of each emitted gluon. Specifically, for
each ordered pair $(ij)$ we compute an energy $E_{ij}$ according to
the distribution
\begin{align}
\dd P_{ij}(E) = &   \frac{\dd E_{ij}}{E_{ij}} \xi_{ij} \Omega_{ij}  
\exp\Big( - \ln \frac{E}{E_{ij}} \xi_{ij} \Omega_{ij} \Big), 
  \label{eq:competition}
\end{align}
where 
\begin{align}
\Omega_{ij} = \frac{\alpha_s}{2\pi} \ln \frac{n_i \cdot n_j}{\lambda}
\end{align}
and
\begin{align}
\xi_{ij} = \sum_{\rho', \rbar'} \frac{\braket{\rho'
    |\rbar'}}{\braket{\tau | \tbar}} \Big| \dbra{\rho'} \bT_i
\ket{\tau} \bra{\tbar} \bT_j \dket{\rbar'} \Big|.
\end{align}
The energy of the previous emission is $E$ and that of the current
emission is $E' = \text{max}(\{ E_{ij} \}, \mu)$. The colour flows
prior to the emission are $\tau$ and $\tbar$. The corresponding pair
$(ij)$ are the parent partons of the emission, the knowledge of which
allows us to select a direction for the emitted gluon. The colour
flows after the emission are selected from the distribution
\begin{align}
P(\rho, \rbar) = \frac{1}{\xi_{ij}} \frac{\braket{\rho
    |\rbar}}{\braket{\tau | \tbar}} \Big| \dbra{\rho} \bT_i \ket{\tau}
\bra{\tbar} \bT_j \dket{\rbar} \Big|.
\end{align}
This choice of $\xi_{ij}$ helps steer the evolution along the most
important trajectories in colour space.
\vspace*{1ex}

{\it Results} -- We consider the production of either a $q \bar{q}$
pair or a pair of gluons ($gg$), with total energy $2Q$ in their zero
momentum frame (ZMF). We refer to these as $V \to q \bar{q}$
(production off a colour singlet gauge boson) and $H \to gg$ (Higgs
decay to gluons) though the cross sections we present are independent
of the details of the initial state, so long as it is a colour
singlet. One interesting observable that is sensitive to wide-angle,
soft-gluon production in events with two primary jets is the `jet
veto' cross section, in which one vetoes events that have one or more
particles radiated into some fixed angular region with energy greater
than some value, $Q_0$. Like most non-global observables, the
summation of even the leading logarithms with full colour accuracy has
not yet been achieved \footnote{One notable exception being the
  numerical summation of the hemisphere jet mass in $e^+e^-$
  collisions in \cite{Hagiwara:2015bia}}. We will show results using
CVolver for events with a veto on particle production in the central
region ($-\pi/4 < \theta < \pi/4$) in the ZMF of the primary two-jet
system. For $q \bar{q}$ production we take $\ket{\mathcal{M}} =
\ket{1}$, corresponding to a single colour connection, and for $gg$
production we take $\ket{\mathcal{M}} = \ket{21} - \ket{12}/N_c$. We
denote the corresponding veto cross section $\Sigma(\rho)$ where $\rho
= Q_0/Q$, i.e. the inclusive cross section is $\Sigma(1) = \text{Tr}
\mathbf{H} = N_c$ for $q \bar{q}$ production and $\Sigma(1) = N_c^2 -
1$ for $gg$ production.

\begin{figure}
\centering
\begin{minipage}{0.35\textwidth}
\flushright
\subfloat[]{\includegraphics[scale=0.75]{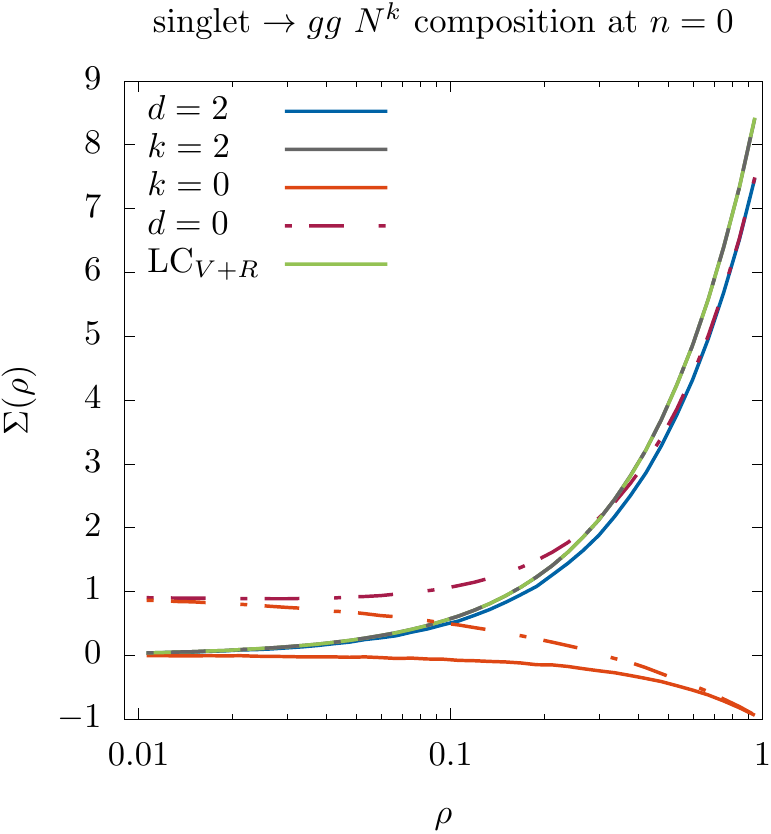}}\\
\subfloat[]{\includegraphics[scale=0.75]{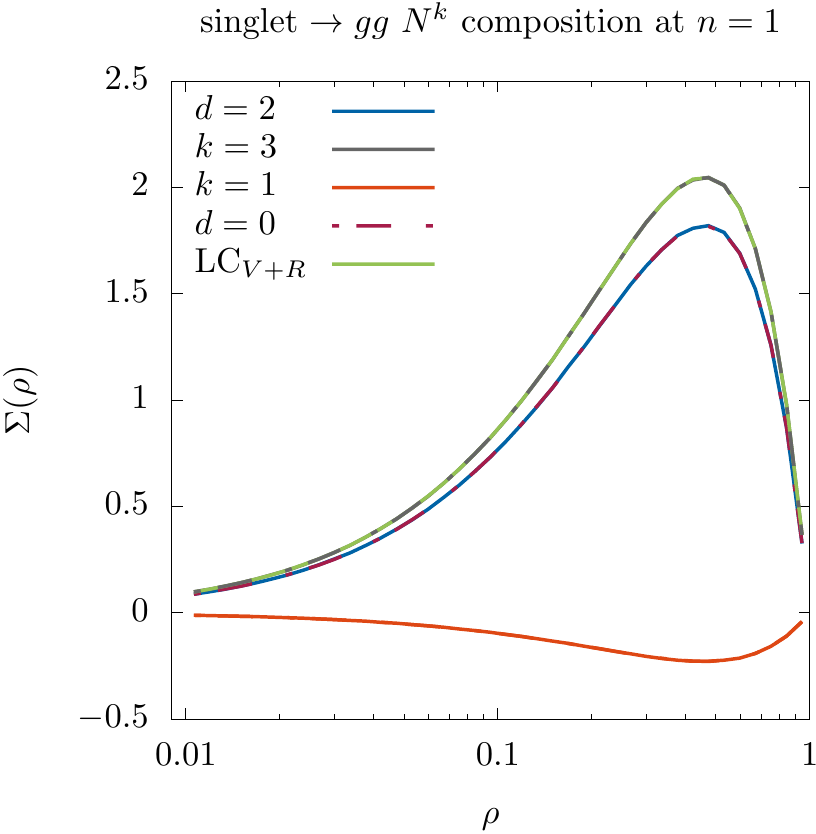}}
\end{minipage}
\caption{\label{fig:ggN1} The jet veto cross section in $H \to gg$ for
  different gluon multiplicities ($n = 0,1$). Also shown are the
  individual contributions at each order of $N_c^k$ for $d=0$ and
  $d=2$. Solid curves (except that labelled LC${}_{R+V}$) always
  correspond to $d=2$ and the broken curves always correspond to
  $d=0$.  See text for details.}
\end{figure}

\begin{figure}
\centering
\begin{minipage}{0.35\textwidth}
\flushright
\subfloat[]{\includegraphics[scale=0.75]{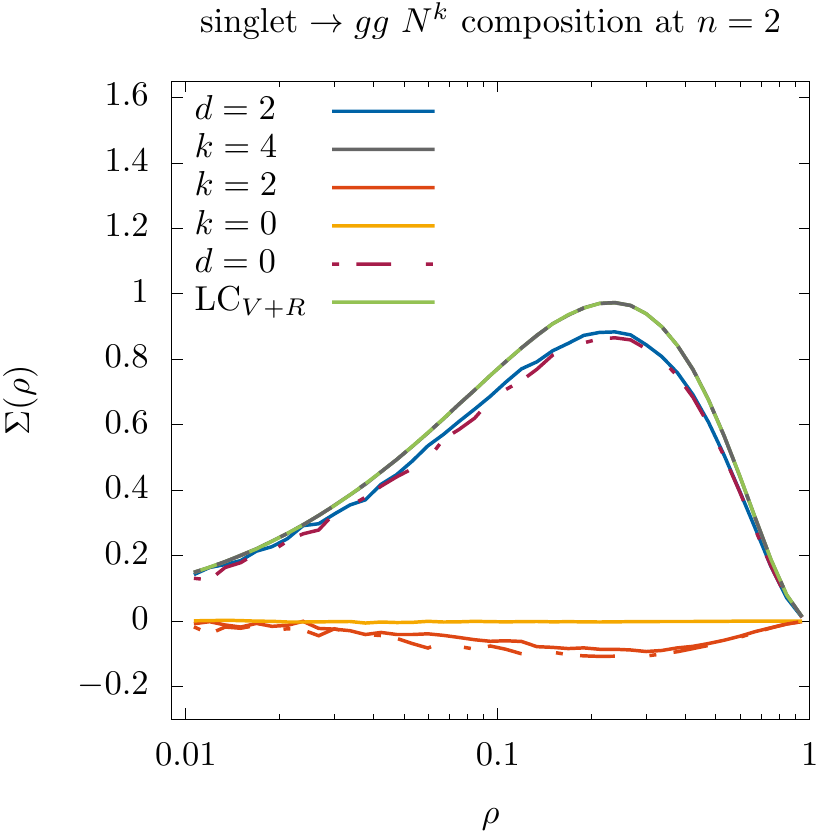}}\\
\subfloat[]{\includegraphics[scale=0.75]{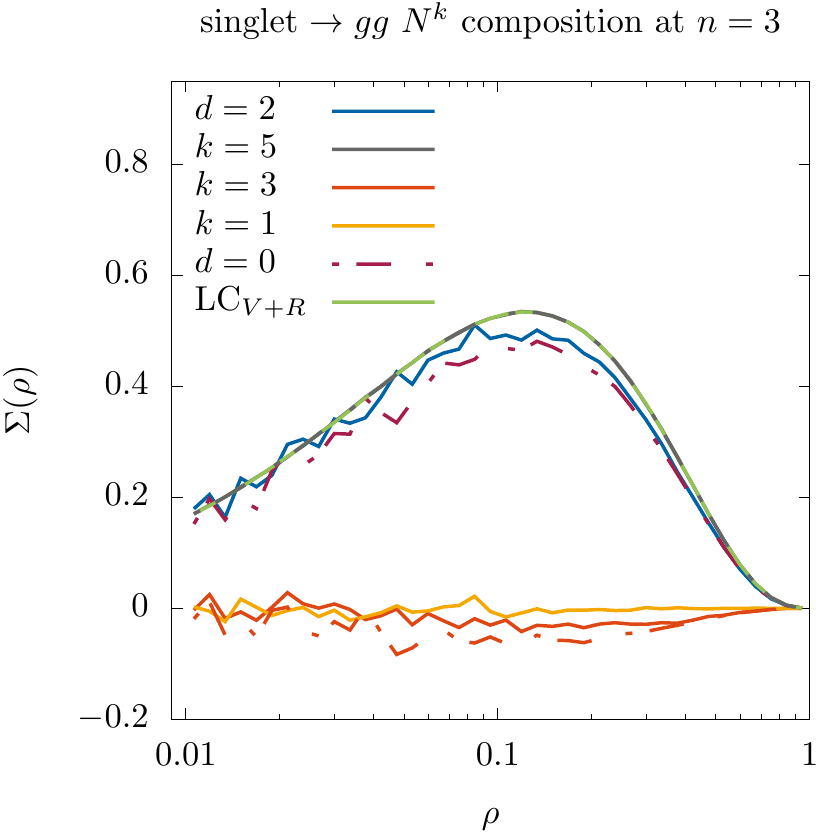}}
\end{minipage}
\caption{\label{fig:ggN2} The jet veto cross section in $H \to gg$ for
  different gluon multiplicities ($n = 2,3$). Also shown are the
  individual contributions at each order of $N_c^k$ for $d=0$ and
  $d=2$. See text for details.}
\end{figure}

Figures~\ref{fig:ggN1} and \ref{fig:ggN2} show the veto cross section
dependence on $\rho$ for different gluon multiplicities up to $n=3$ in
$H \to gg$. We present results for \mbox{$d=0$} (LC${}_V$) and $d=2$
(NNLC${}_V$). In the cases we consider here the differences between
\mbox{$d=1$} and $d=2$ are always less than $1-2$\% for up to two
emissions or otherwise well within the fluctuations we observe. Also
shown are the strictly leading-colour results, LC$_{V+R}$, which are
composed of the $N_c^{k_n^{\text{max}}}$ contribution in
Eq.~\eqref{eq:nbreakdown} below, with $\Sigma_n^{(k_n^{\text{max}})}$
calculated in the $d=0$ approximation. Finally, we show the breakdown
in terms of the different powers of $N_c$ that
contribute. Specifically, we consider
\begin{align}
    \Sigma_n = \sum_k^{k_n^{\text{max}}} N_c^k \Sigma_n^{(k)},
    \label{eq:nbreakdown}
\end{align}
where $k_n^{\text{max}} = n+1$ for $q \bar{q}$ production and
$k_n^{\text{max}} = n+2$ for $gg$. The coefficients $\Sigma_n^{(k)}$
include Sudakov $R$-factors, such as Eq.\eqref{eq:R1} and
Eq.\eqref{eq:R2}, whose exponents are given by the diagonal entries in
the anomalous dimension matrix. This is not a strict expansion in
powers of $N_c$. Rather it keeps track of $1/N_c$, off-diagonal
suppression that occurs in the hard scattering matrix, the scalar
product matrix $(\braket{\alpha | \beta})$, the real emission
operators and the successive terms (indexed by $l$) in
Eq.\eqref{eq:simon}.  The case $n=0$ illustrates a peculiar feature of
our LC${}_{V}$ approximation for $H \to gg$: the non-vanishing at
small $\rho$ occurs because of an unphysical $N_c^0$ contribution that
is present at $d=0$ due to the subleading $N_c$ terms in the scalar
product and hard scattering matrices: the strictly leading $N_c$
contribution corresponds to the well-behaved LC${}_{V+R}$ curve, which
is equal to the $k=2$ curve and is the same for $d=0$ and $d=2$.

\begin{figure}
\centering
\begin{minipage}{0.35\textwidth}
\flushright
\subfloat[]{\includegraphics[scale=0.75]{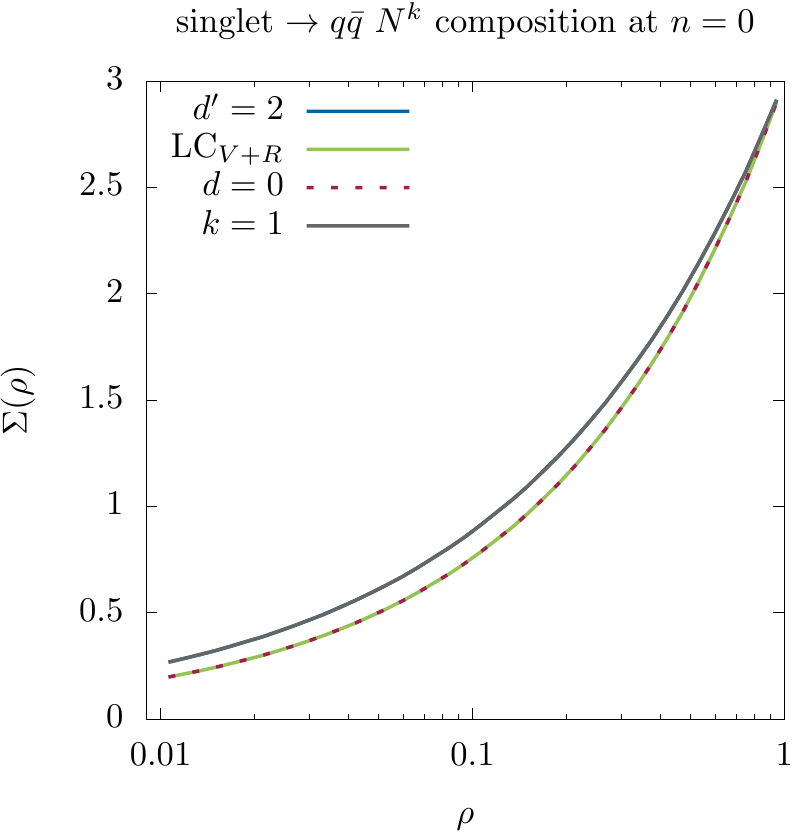}}\\
\subfloat[]{\includegraphics[scale=0.75]{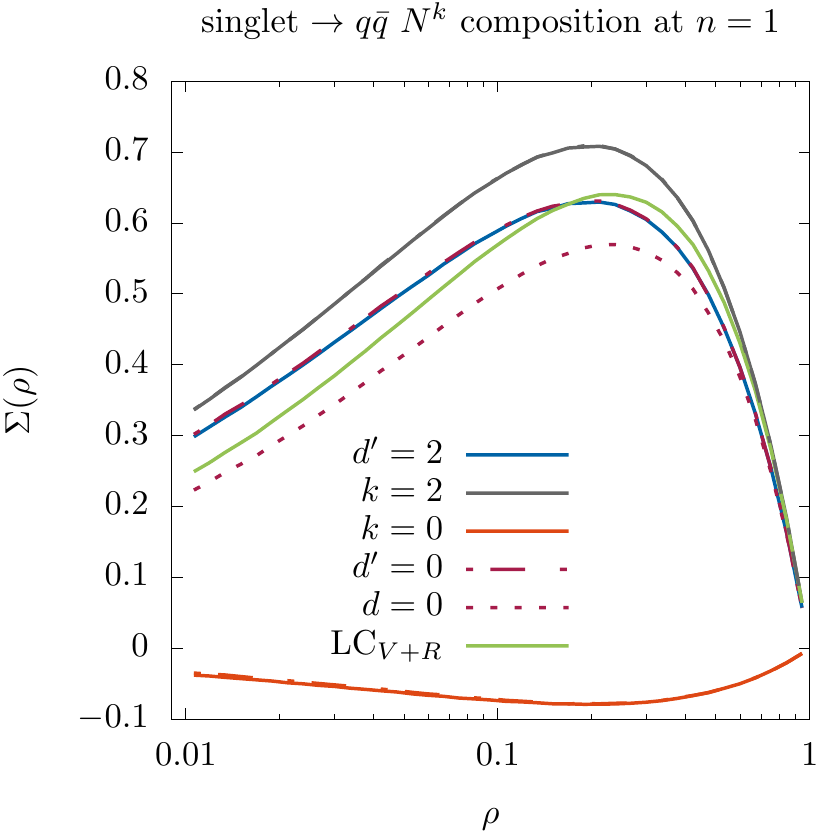}}
\end{minipage}
\caption{\label{fig:qqN1} The jet veto cross section in $V \to q
  \bar{q}$ for different gluon multiplicities ($n = 0,1$). Also shown
  are the individual contributions at each order of $N_c^k$ for $d=0$
  and $d'=2$. See text for details.}
\end{figure}

\begin{figure}
\subfloat[]{\includegraphics[width=0.35\textwidth]{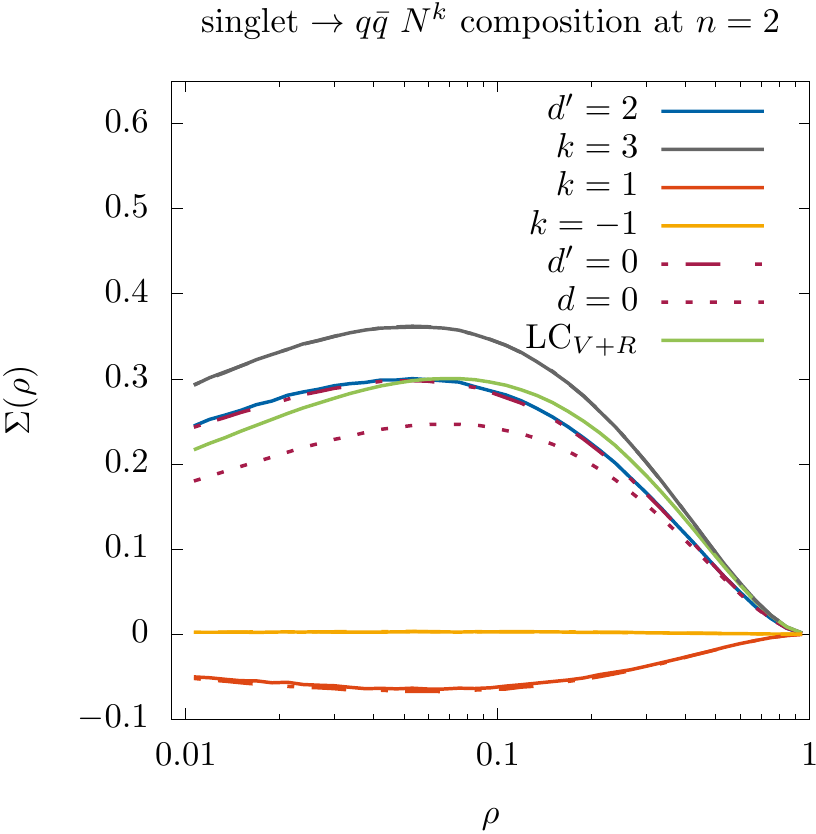}}\\
\subfloat[]{\includegraphics[width=0.35\textwidth]{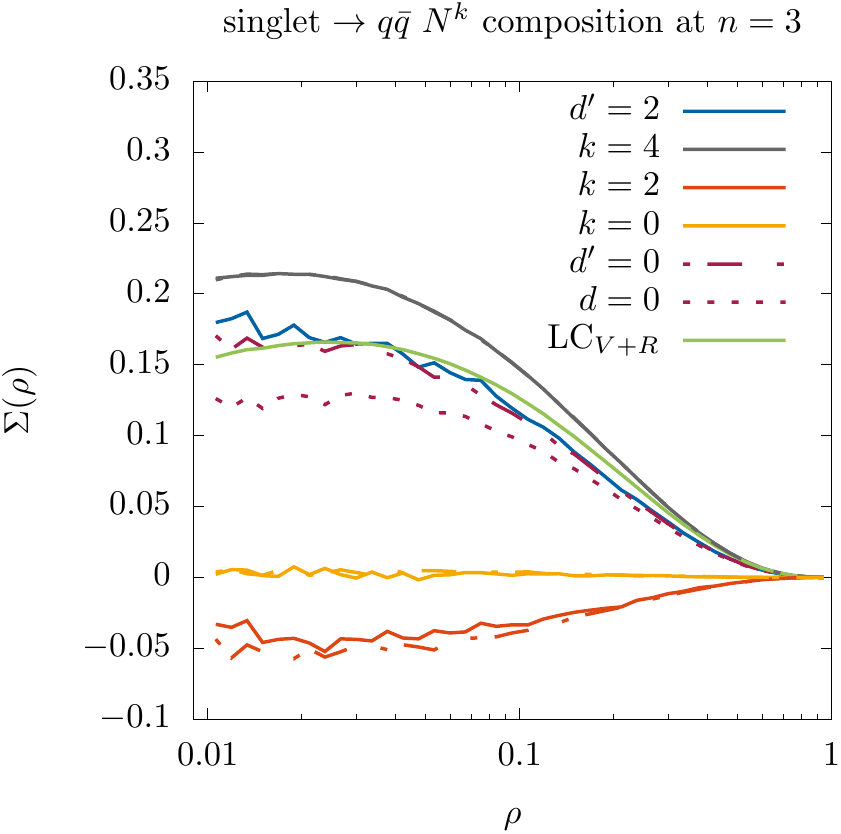}}\\
\caption{\label{fig:qqN2} The jet veto cross section in $V \to q
  \bar{q}$ for different gluon multiplicities ($n = 2,3$). Also shown
  are the individual contributions at each order of $N_c^k$ for $d=0$,
  $d^{\prime}=0$ and $d'=2$. See text for details.}
\end{figure}

\begin{figure}
\centering
\begin{minipage}{0.35\textwidth}
\flushright
	\subfloat[]{\includegraphics[scale=0.75]{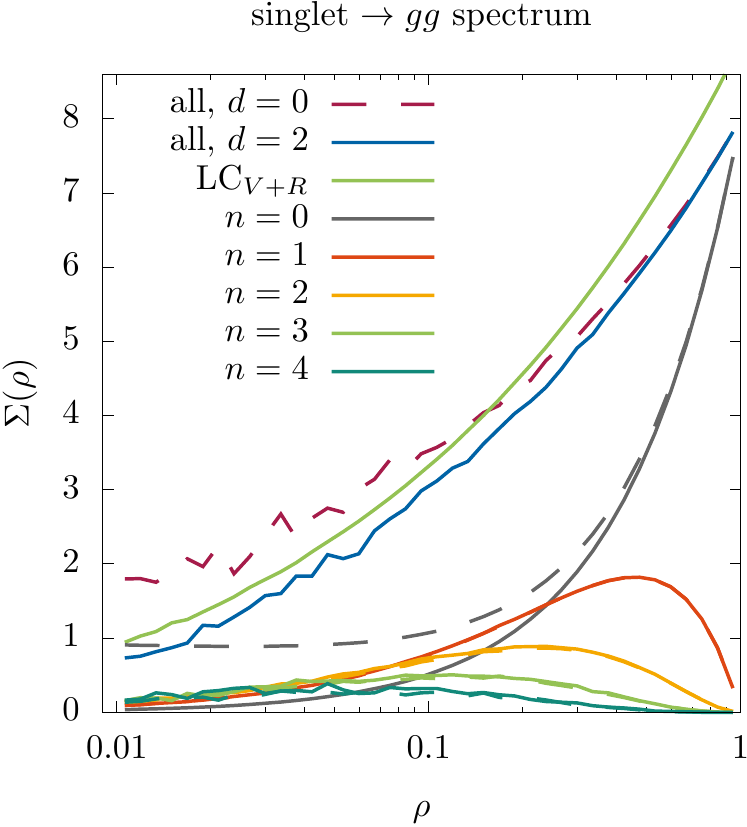}} \\
	\subfloat[]{\includegraphics[scale=0.75]{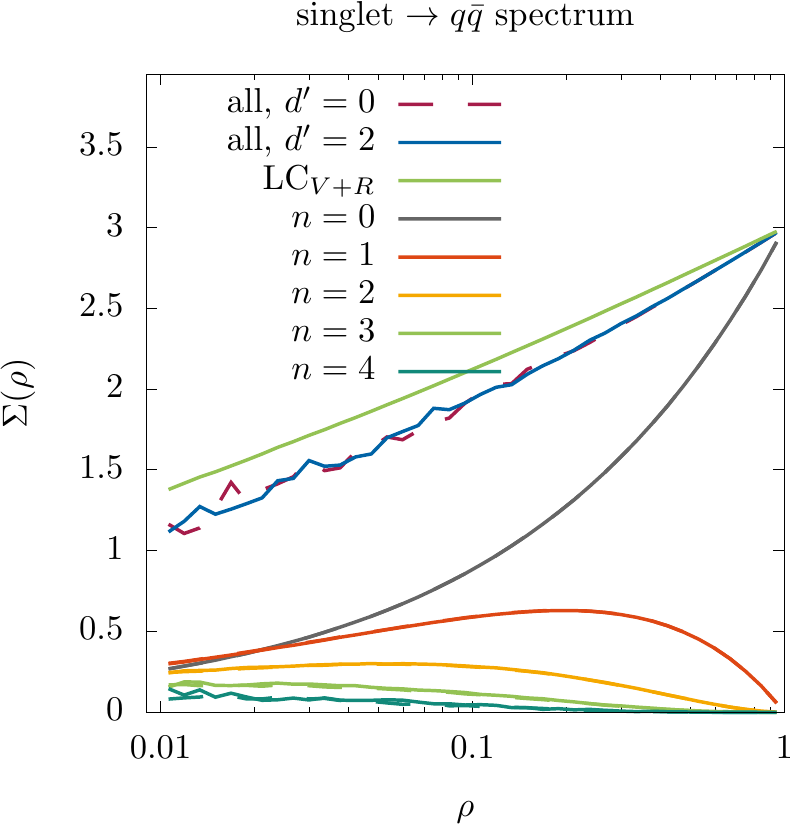}}
	\end{minipage}
	\caption{\label{fig:full} The jet veto cross section in (a) $H
          \to gg$ and (b) $V \to q \bar{q}$ for different gluon
          multiplicities. Results are shown for $d=0$ and $d=2$. We
          show the contributions from $n=0$ up to $4$ emissions,
          however the complete result sums over all emissions (in
          practice we have limited the simulation to $n\le 40$
          emissions).}
\end{figure}

 Figures~\ref{fig:qqN1} and \ref{fig:qqN2}, are the corresponding
 plots for the $V \to q \bar{q}$ process. Here we use the notation $d$
 and $d^{\prime}$ to distinguish between exponentiating and not
 exponentiating the colour-diagonal $\rho$-term in the anomalous
 dimension matrix. Note that for $n = 0$, the $d'=0$ result is the
 exact result for quarks and the $n=1$, $d=0$ result is the exact
 result for gluons.  Finally, in Fig.~\ref{fig:full}, we show how the
 total cross section is built up from different multiplicities. The
 differences between the full-colour (solid blue) curves and the
 leading colour (LC${}_{V+R}$) ones are clearly significant. The
 apparent success of our $d'=0$ approximation is interesting to note,
 though we do not expect this to continue once we consider more
 sophisticated hard processes.

{\it Conclusions} -- This paper represents a major milestone in a
project to compute numerically full-colour evolution in perturbative
QCD. The inclusion of subleading colour effects will improve the
accuracy of future simulation codes and, as a result, be of
considerable value to experimenters and theorists interested in
performing percent-level simulations for current and future colliders.
For the future, we intend to go beyond the soft approximation and
include incoming hadrons.
\vspace*{1ex}

{\it Acknowledgements} -- The authors want to thank the Erwin
Schr\"odinger Institute Vienna for support while this work has been
finalized.  JRF thanks the Institute for Particle Physics
Phenomenology in Durham for the award of an Associateship. This work
has received funding from the UK Science and Technology Facilities
Council, the European Union’s Horizon 2020 research and innovation
programme as part of the Marie Skłodowska-Curie Innovative Training
Network MCnetITN3 (grant agreement no. 722104), and in part by the
COST actions CA16201 ``PARTICLEFACE'' and CA16108 ``VBSCAN''. We are
grateful to Thomas Becher, Jack Holguin, Mike Seymour, Malin
Sj\"odahl, and Ren\'e \'Angeles Mart\'inez for discussions.

\bibliography{refs}

\end{document}